\documentclass[twocolumn,superscriptaddress,prl,floatfix,showpacs]{revtex4}
\usepackage[]{graphicx}
\usepackage[usenames,dvipsnames]{color}
\usepackage{soul}
\usepackage{bm}

\begin{document}

\title{Intertwined Rashba, Dirac and Weyl Fermions in Hexagonal Hyperferroelectrics}

\author{Domenico Di Sante} 
\affiliation{Consiglio Nazionale delle Ricerche (CNR-SPIN), Via Vetoio, L'Aquila, Italy}\email{domenico.disante@spin.cnr.it}
\affiliation{Department of Physical and Chemical Sciences, University of L'Aquila, Via Vetoio 10, I-67010 L'Aquila, Italy}

\author{Paolo Barone}
\affiliation{Consiglio Nazionale delle Ricerche (CNR-SPIN), Via Vetoio, L'Aquila, Italy} 

\author{Alessandro Stroppa} 
\affiliation{Consiglio Nazionale delle Ricerche (CNR-SPIN), Via Vetoio, L'Aquila, Italy}
%\email{alessandro.stroppa@aquila.infn.it}

\author{Kevin F. Garrity}
\affiliation{Material Measurement Laboratory, National Institute of Standards and Technology, Gaithersburg MD, 20899, USA}

\author{David Vanderbilt}
\affiliation{Department of Physics and Astronomy, Rutgers University, Piscataway, New Jersey 08854, USA}

\author{Silvia Picozzi}
\affiliation{Consiglio Nazionale delle Ricerche (CNR-SPIN), Via Vetoio, L'Aquila, Italy}

\date{\today}

\begin{abstract}

By means of density functional theory based calculations, we study  the
role of spin-orbit coupling in the new family of $ABC$ hyperferroelectrics [Phys. Rev. Lett. 112, 127601 (2014)].
We unveil an extremely rich physics strongly 
linked to ferroelectric properties, ranging from 
the electric control of bulk Rashba effect to the existence of a three dimensional topological insulator phase,
with concomitant topological surface states even in the ultrathin film limit. Moreover,
we predict that the topological transition, as induced by alloying, is followed by a 
Weyl semi-metal phase of finite concentration extension, which is robust against disorder, putting forward hyperferroelectrics
as promising candidates for spin-orbitronic applications.
  
\end{abstract}


\maketitle

{\it Introduction.} Spin-orbit coupling (SOC) is a relativistic interaction that
gives rise to a rich variety of interesting
phenomena in solid
state physics, ranging from topological quantum phases of matter to
Rashba/Dresselhaus-like spin splitting effects.
In the former case, fully
spin-polarized massless Dirac fermions appear at the surface of  topological
insulators (TIs), showing  protected metallic surface states
despite their bulk insulating character \cite{TI1,TI2}. In the latter case, which
is realized at surfaces or interfaces as due to a structural
inversion asymmetry of
the confinment potential or in non-centrosymmetric materials,
spin-splitting effects on the  bands of massive fermions appear,
which may ultimately give rise to a spin-Hall effect. \cite{SH-dyakonov, Rashba1984} 
The presence of strong spin-momentum locking in both systems allows, in
principle, for an all-electric control of the electron spins, putting forward
these material systems as ideal candidates for new spintronic
applications.

Interestingly, SnTe, the first
proposed Topological Crystalline Insulator (TCI) in its centrosymmetric
structure \cite{SnTeNatComm}, was predicted to host a  coexistence of
Rashba and topological properties when the inversion symmetry is broken at  low
temperature by a ferroelectric instability. \cite{SnTePRB} In fact, 
the breaking of inversion symmetry alone is not expected to
suppress the topological insulating phase, as long as the relevant symmetries
responsible of  the topological protection are maintained. On the other
hand, the gapless Dirac states at the surface of an acentric TI are expected to
have completely different shapes on inequivalent sides of the material; as a
consequence, electrons would be fractionalized into nonequivalent halves on
opposite surfaces (showing, e.g., charge carriers of different $p$/$n$ types or
parallel spin-polarization pattern), possibly leading to new features and
spintronic functionalities. \cite{prl_Moore2010,top_pn_2012, BiTeI3}
Recently it has been shown that the phase transition from a 
noncentrosymmetric three-dimensional topological
insulator to a normal insulator always hosts an
intermediate Weyl semimetal phase. \cite{savrasov_prb2011, vanderbilt_prb2014, Murakami1, Murakami2}
The interplay of the above features  
with ferroelectricity has not been
considered so far, despite the fact that ferroelectrics, {\it i.e.}, polar materials with
switchable electric polarization, 
could provide an additional degree of freedom  for 
controlling  and/or manipulating the electronic
properties of relativistic fermions. 
A very large Rashba spin-splitting in the bulk electronic structure
has been reported for tellurohalides, a family of noncentrosymmetric, but not ferroelectric,
semiconductors with strong SOC. \cite{BiTeI2,BiTeI1,BiTeCl,Tellurohalides}
After that, it has been shown that ferroelectric polarization may allow  
the permanent  control of bulk Rashba-like effects, 
such as the switching of spin-texture upon reversal of 
ferroelectric polarization,  
thus allowing for new non-volatile 
functionalities in spintronic devices.
GeTe, one of the oldest
ferroelectrics, has been considered as the prototypical Ferroelectric Rashba
Semiconductor (FERSC). \cite{GeTe,FERSC,GeTeEXP}

Most of the well-known conventional ferroelectrics belong to the class of
insulating transition-metal perovskite oxides. In these systems, 
the ferroelectric transition arises typically from an unstable zone-centered transverse optic
(TO) mode associated with a symmetry-lowering polar distortion of the bulk
material. \cite{LOTOVanderbilt}
However, a thin film of such (proper) ferroelectric materials will experience
large depolarization fields arising from surface charge accumulations, which
would produce a strong electric field counteracting the polar displacements. It 
has been shown that the depolarization field   
completely suppresses the ferroelectric distortion in films thinner
than $\sim\,10~nm$. \cite{ghosez_nature2003} This issue, together  with their
strong insulating character, make conventional proper ferroelectrics unsuitable
for real technological applications aiming at integrating ferroelectricity and
relativistic effects in low-dimensional devices.
However, very recently, a new class of ferroelectric materials,
called "hyperferroelectrics", has been  proposed. \cite{Hyperferro} The
interesting property of these systems is their ability to spontaneously polarize
even in the presence of an unscreened depolarization field. This feature arises
from an additional instability of a longitudinal optic (LO) mode besides the
usual TO mode instability characteristic of proper
ferroelectrics. \cite{Hyperferro} Garrity and coworkers proposed that hexagonal
$ABC$ ferroelectric semiconductors with LiGaGe structure
are suitable candidates for hyperferroelectricity. \cite{ABCLiGaGe,Supp} 

In this study, we focus on the relativistic properties
of these materials by means of accurate
density functional theory (DFT) calculations. \cite{Supp} We show that hexagonal
hyperferroelectrics (HHs) belong to and broaden the class of FERSC materials,
where the bulk Rashba spin-texture can be fully controllable by the
ferroelectric polarization. Since  hyperferroelectrics
are less sensitive to surface effects, the switchable
bulk Rashba features are expected to survive in low-dimensional systems, which 
is clearly an appealing property for spintronic applications. 
Furthermore, we show that HHs can
host a topologically non trivial $\mathcal Z_2$ quantum phase with
symmetry-protected {\it inequivalent and tunable} surface states coexisting with
Rashba-like features. While the topological phase is predicted to survive up to
ultrathin films, we also show that a phase transition from topological insulator
to a Weyl semimetal and eventually to a normal insulator can be realized by
chemical doping and alloying.

\begin{figure}[!t]
\centering
\includegraphics[width=\columnwidth,angle=0,clip=true]{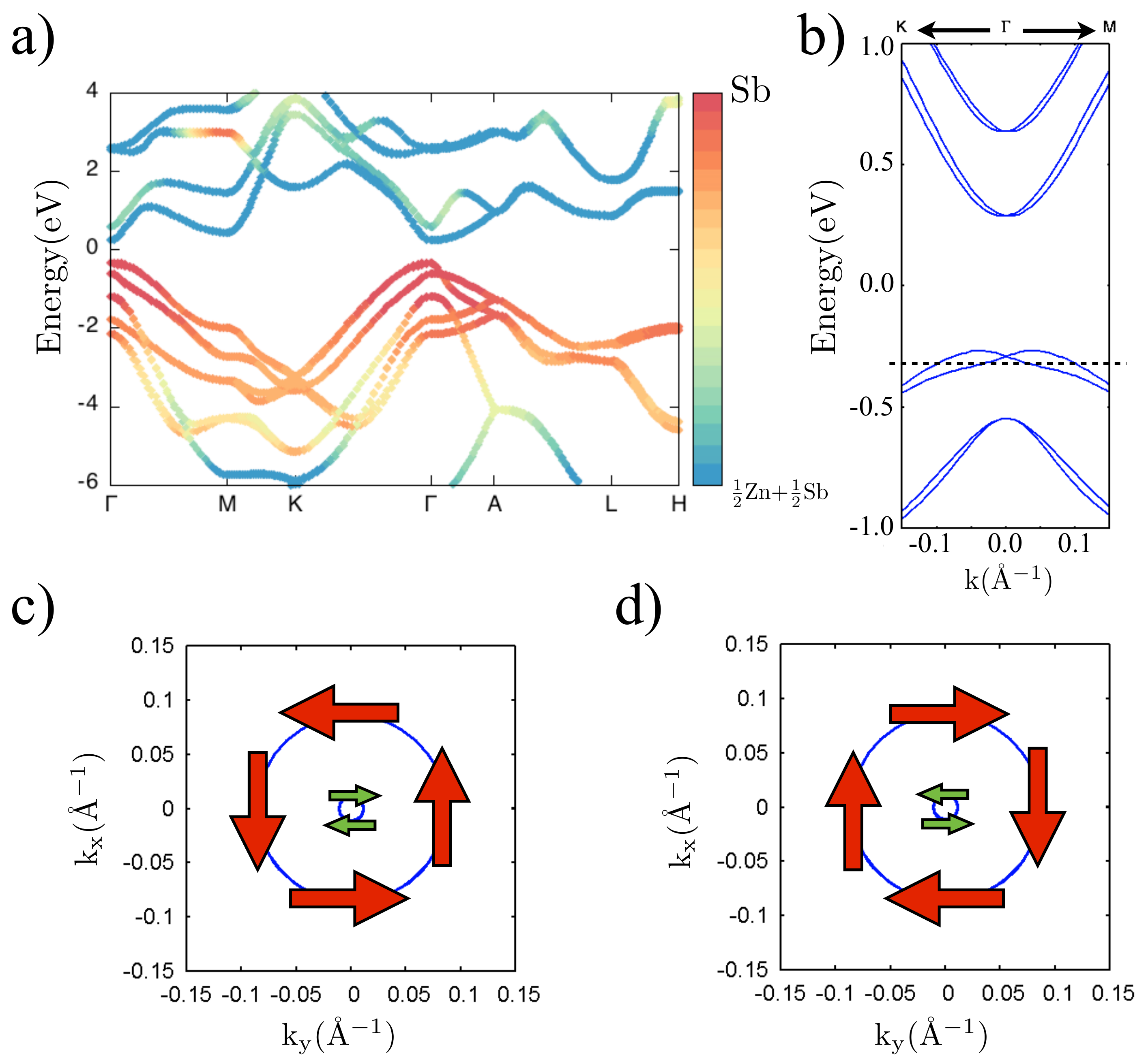}
\caption{(Color online) a) Relativistic bandstructure of NaZnSb along the
high-symmetry lines of the hexagonal Brillouin zone.\cite{Supp}
Colors highlight the atomic orbital character, from Sb orbitals (red) to an
equal mix of Zn and Sb orbitals (blue). b) Zoom at the $\Gamma$ point around
the Fermi level along the $\Gamma$K and $\Gamma$M directions. A Rashba type
spin-splitting occurs at the valence, as well as conduction, bands. c)
Spin-texture of the valence bands at -0.30 eV (dashed line in panel b), with
spins circularly rotating on both branches. d) Spin-texture reversing after the
switching of the bulk polarization.}
\label{fig2}
\end{figure}

{\it Results and discussions.---} We begin our analysis by calculating the
relativistic bulk band structure of the hyperferroelectric NaZnSb. In
Fig. \ref{fig2}a), a direct gap is clearly visible at the $\Gamma$ point, where the Sb
orbital character dominates the valence low energy states, while an equally
mixed Sb and Zn character is found in the conduction bands. The expected
dispersion of a Kramers doublet around the $\Gamma$ point can be generally deduced
from the knowledge of the symmetry properties of the corresponding group of $k$.
Since all considered hexagonal $ABC$ ferroelectric semiconductors belong to
space group $P6_3mc$ (no.186), the little group of $\Gamma$ is $C_{6v}$, consisting of
two-fold $C_2$, three-fold $C_3$ and six-fold $C_6$ rotations along the $z$
direction and of three mirror operations $M_v$ about a vertical reflection plane
(containing the $\Gamma$M line) and three $M_d$ about a dihedral reflection
plane (containing the $\Gamma$K line). For spin 1/2 electrons, $C_n$ rotations
can be represented as $e^{-i\sigma_z\pi/n}$, and the two mirror operations as
$M_v=i\sigma_x$ and $M_d=i\sigma_y$, where $\sigma_{x,y,z}$ are the Pauli
matrices for spin degrees of freedom. Imposing the invariance of the $\bm
k\cdot\bm p$ Hamiltonian around $\Gamma$ with respect to the crystalline
symmetry and time-reversal symmetry $T=i\sigma_y \mathcal{K}$ (where
$\mathcal{K}$ is complex conjugation), one finds
$H_{\Gamma}=(k_x^2+k_y^2)/2m_{xy}^*+k_z^2/2m_z^*+\alpha_R(k_x\sigma_y-k_y\sigma_x)$, 
where the last term describes a Rashba coupling, while $m_{xy}^*,
m_z^*$ are the in-plane and out-of-plane effective masses. Third-order terms
leading to warping of the bands are symmetry-forbidden, since they
would break the invariance with respect to $C_2$ and $C_6$ rotations. A zoom
around the $\Gamma$ point highlights typical Rashba-like spin-split bands, both
in the valence and conduction regions, as shown in Fig.~\ref{fig2}b) along the two
orthogonal $k_x$ ($\Gamma$K) and $k_y$ ($\Gamma$M) axes. 

\begin{figure}[!ht]
\centering
\includegraphics[scale=0.20,angle=0,clip=true]{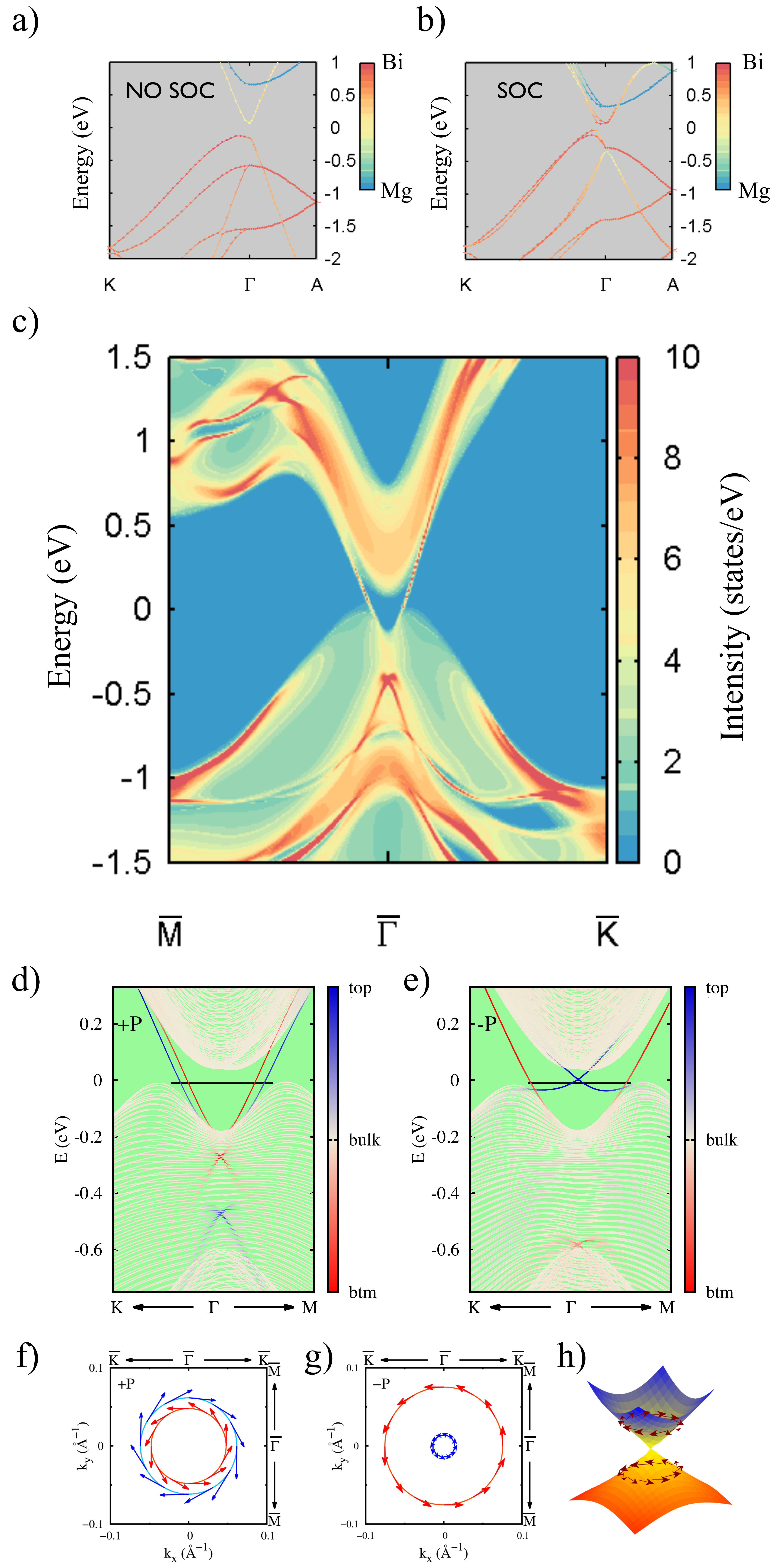}
\caption{(Color online) Zoom at the $\Gamma$ point around the Fermi level
without (panel a) and with (panel b) SOC for KMgBi. 
c) Surface Spectral Function for the
(0001) hexagonal surface Brilouin zone where topological
non-trivial surface states are visible inside the bulk bandgap.\cite{Supp}
d) and e) Surface states of KMgBi calculated in slab geometry for
opposite direction of ferroelectric polarization.
Characters of top (MgBi-terminated) and bottom (K-terminated) surface states are
highlighted by the color scale, revealing that the Dirac cones of K-terminated
surfaces are always buried in the continuum of bulk states. f), g) display spin
textures of the surface states at the energy cut shown in panels d), e), while
h) shows that the chirality of spin polarization is reversed when crossing the
Dirac cone.}
\label{fig3}
\end{figure}

An energy cut below the degeneracy point at $\Gamma$ reveals an isotropic Fermi
surface in the $k_xk_y$-plane, with two concentric circular branches (see Fig.
\ref{fig2}c). Spins, always tangential to the isoenergy curves and perpendicular
to momenta, rotate in opposite ways in the two branches, consistently with the
$\bm k\cdot\bm p$-model sketched above.
%$\alpha_R(k_x\sigma_y - k_y\sigma_x)$. Warping effects due to cubic terms in the
%Rashba Hamiltonian, important in GeTe and tellurohalides\cite{GeTe,BiTeI1}, are
%negligible in NaZnSb.\cite{GeTe,BiTeI1} 
We can estimate the strength of the Rashba effect through the energy and
momentum offsets $E_R$ and $k_R$ locating the band extrema via the Rashba
parameter $\alpha_R=2E_R/k_R$.
For NaZnSb we find $k_R\approx0.038$\,\AA$^{-1}$,
$E_R\approx\,42$\,meV and $\alpha_R\approx 1.09$\,eV\AA\ in the valence band and
$k_R\approx0.004$\,\AA$^{-1}$, $E_R\approx 4$\,meV and $\alpha_R\approx
0.32$\,eV\AA\ in the conduction band, which are comparable to, if not larger
than, typical values estimated at the surfaces of heavy metals such as Bi or
Au. The FERSC behavior emerges when the ferroelectric polarization is reversed,
leading to a complete switching of the spin texture, as it is clear by comparing
Fig. \ref{fig2}c) and \ref{fig2}d). The switching can be undestood as arising
from a change of sign of the Rashba parameter $\alpha_R$, which appears to
depend linearly on the polar displacements (in analogy with the GeTe case \cite{GeTe}).
Therefore, hyperferroelectrics can be considered as new multifunctional
materials, where the control of the electric degrees of freedom has a direct
consequence on the spin properties.

\begin{figure*}[!t]
\centering
\includegraphics[width=\textwidth,angle=0,clip=true]{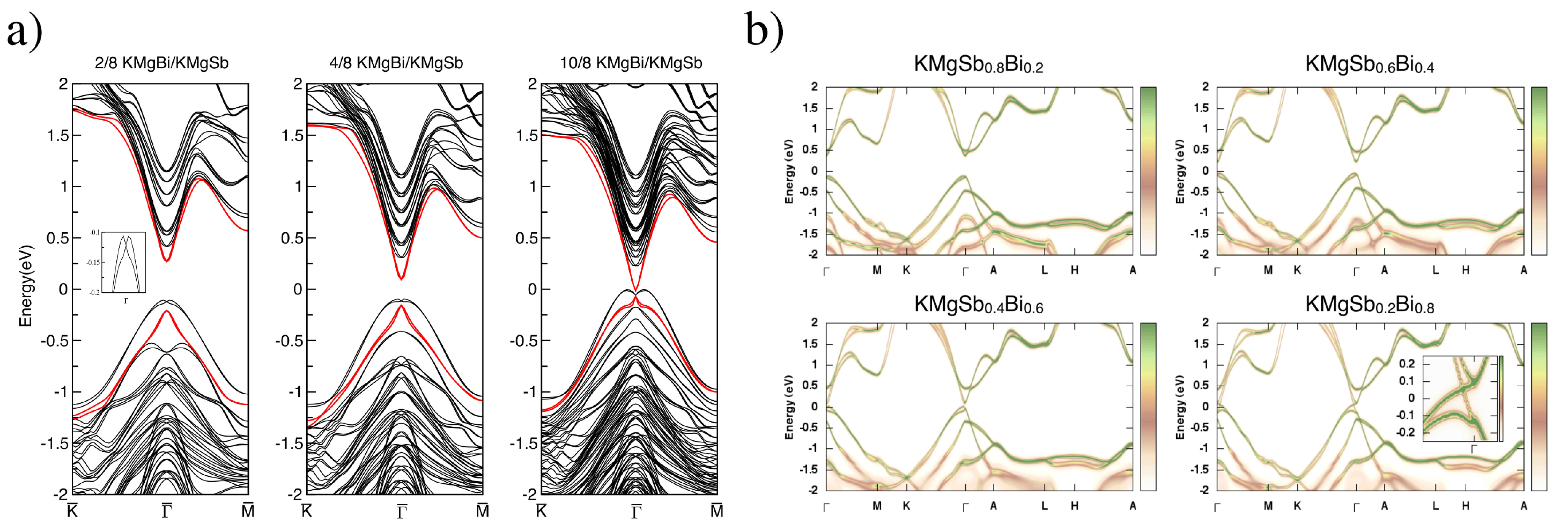}
\caption{(Color online) a) Bandstructures for 2/8, 4/8 and 10/8 KMgBi/KMgSb
superlattice configurations along the high-symmetry directions $\bar{\text K}$-$\bar{\Gamma}$-$\bar{\text M}$. 
Red lines emphasise topologically non-trivial interfacial states emerging from
the continuum of bulk bands (black lines), while the inset is a zoom of the valence Rashba bands. 
b) Spectral functions for diluite 
KMgSb$_{1-x}$Bi$_x$ alloys at $x=0.2$, $0.4$, $0.6$ and $0.8$ respectively.\cite{Supp}
In the last panel, the inset reports a zoom around the Fermi level highlighting
the Weyl semi-metal phase. In this energy region, the three dimensional nature 
of the Weyl fermions' dispersion ensures protection against the renormalisation
effects of disorder.\cite{DiSanteSciRep} The finite bandwidth is only due to a finite $\delta=10$ meV
in the Green's function calculation.\cite{Supp}}
\label{fig5}
\end{figure*}

We now consider KMgBi. The  bandstructure is
similar to that shown in Fig. \ref{fig2}a) for NaZnSb \cite{Supp}, with Rashba-like
spin-split bands around the $\Gamma$ point. The
valence bands present a dominant Bi
orbital character, while a mixture of Bi and Mg character is present in the
conduction states. As for the case of NaZnSb, KMgBi belongs to the family of
FERSCs, since the Rashba spin texture can be switched by switching the
ferroelectric polarization. A crucial difference between the two compounds
emerges when inspecting the orbital character around the band gap at the
$\Gamma$ point. Comparing Fig. \ref{fig3}a) with \ref{fig3}b), we note that the
presence of SOC leads to an inverted gap where anionic (Bi) and cationic (Mg)
characters are exchanged, while leaving an indirect band gap of about $0.1$ eV.
To assess the topological nature of KMgBi, we calculated the topological indexes
following the adiabatic pumping of the Wannier charge
centers \cite{TOPOinv,TOPOinv2},
finding that KMgBi is a strong 3D topological insulator
with (1;000) indices. \cite{Supp} 

The first consequence of a non trivial gap is the appearance of metallic
massless surface states when the system is terminated by vacuum
(semi-infinite slab with Mg-Bi termination)
as shown in Fig.~\ref{fig3}c),
where the Dirac point appears to be buried in the continuum of
bulk states. 
Remarkably, the properties of the surface states appear to  depend strongly on the
nature of the termination and on the direction of the ferroelectric
polarization. In Figs.~\ref{fig3}d) and e), we show the calculated band structure for two
opposite directions of ferroelectric polarization in the slab
geometry, which allows us to consider the surface states of the inequivalent
terminations in a single calculation. In this geometry, the bottom
termination always comprises K ions, while the top one consists of Mg/Bi ions.
Here we note that the slab calculations in Fig.~\ref{fig3} have been performed 
within the tight-binding framework to verify the bulk-edge correspondence, while neglecting possible
self-consistent charge rearrangments at the surfaces due to the presence of
polar terminations (K$^{+}$ and [MgBi]$^{-}$). \cite{Supp}
Surface states arising from different terminations can then be identified by
calculating the ionic character of the bands coming from superficial K or Mg/Bi
ions. It is evident that the Dirac cones are
buried in the bulk bands when the crystal terminates with K ions, even though a
reversal of the ferroelectric polarization causes a significant modulation of
the Dirac point position. On the other hand, at the MgBi termination the Dirac
cone can be moved from the bulk valence continuum into the gap, close to the
conduction bands. Even though the inequivalence of the surface states has been
known as a characteristic feature of noncentrosymmetric TIs, we want to emphasize 
the strong tunability of the Dirac cones arising from the unique
interplay with ferroelectricity. An in-gap energy cut reveals different
Fermi surfaces and spin-textures depending on the polarized state of the bulk
system. For instance, at positive $P$ the charge carriers from surface states would
both show a $n-$type character,
being above the corresponding Dirac cones, and
opposite spin chiralities shown in Fig. \ref{fig3}f). However, when the polarization
is reversed, the charge carriers at the top termination are holes, sharing the
same spin chirality of bottom spin-polarized states (see Fig. \ref{fig3}g)).
Side-by-side domain walls between different ferroelectric domains could behave
then as intrinsic topological $p$-$n$ junctions, with interesting implications for
novel spintronic devices. \cite{top_pn_2012} 
Due to time-reversal symmetry, surface states above and below a Dirac
point would always show opposite spin chirality, as shown in Fig.
\ref{fig3}h).

We recall here that an important  property  of
HHs is their ability to sustain a ferroelectric polarization
down to single atomic layers, even when interfaced with a normal insulator or
confined in a slab geometry terminated by vacuum. \cite{Hyperferro} Therefore,
HHs are suitable systems for studying the 
interplay between ferroelectric and spin properties in the low
dimensional limit. We placed topologically non-trivial KMgBi in superlattice configurations
with thick slabs of topologically trivial KMgSb (hereafter we adopt the same n/m
ABC/A'B'C' notation used in Ref.~\cite{Hyperferro}). 
In Fig. \ref{fig5}a) we report the bandstructures for 2/8, 4/8, 10/8 KMgBi/KMgSb
superlattices, corresponding to about 0.8, 1.6 and 4.0 nm of topological thin films
respectively, embedded in a polar trivial host. Standard ferroelectrics 
(\textit{i.e.}, not hyperferroelectris) will not
polarize in this ultrathin geometry, while we checked that KMgBi remains polar
even for the thinner 2/8 superlattice configuration (with KMgSb both in the 
polar and centrosymmetric phases). A polar structure is confirmed by the  
Rashba-split bands (see inset in Fig. \ref{fig5}a), whose chiralities are electrically controllable by acting on
the ferroelectric polarization. Moreover, the topological nature of KMgBi thin
films leads to the appearance of topologically protected interfacial states (red lines in Fig. \ref{fig5}a), 
becoming gapless when the interaction between the two KMgBi/KMgSb interfaces
vanishes (as for the 10/8 configuration). Unlike for the slab geometry,
where the surfaces are terminated by vacuum, superlattices' interfacial electronic structures
don't show, by construction, a dependence
on the polarization direction, and Dirac cone localized at the top 
and bottom interfaces are degenerate in energy.

A further interesting characteristic of topological (hyper)ferroelectrics is
their capacity to undergo a finite intermediate Weyl semi-metal phase during the
topological transition from three-dimensional topological to normal insulators
as a consequence of the broken inversion symmetry. \cite{vanderbilt_prb2014,Murakami1,Murakami2}
In fact, as shown in Fig.~\ref{fig5}b), tuning the relative Bi/Sb ratio in a
diluite KMgSb$_{1-x}$Bi$_x$ alloy brings the system toward a topological transition
marked by a gap closure.\cite{Supp} In this semi-metal phase of finite concentration extension 
(0.65 $<$ $x$ $<$ 0.90)\cite{Supp}, bands linearly cross
around the Fermi level (see inset in Fig.~\ref{fig5}b for the case of KMgSb$_{0.2}$Bi$_{0.8}$),
giving rise to a Weyl semi-metal phase.\cite{savrasov_prb2011}
It is worth noting that such a phase is protected against the
broadening effects of disorder, as a natural consequence of the three dimensional
nature of its electronic structure.\cite{DiSanteSciRep} 
Furthermore, the distinct power law of the density of states within the Weyl phase (i.e., $D(\omega)\propto \omega^2$)
leads to a change of exponents in the temperature dependence of thermodynamic quantities
such as the specific heat ($C_V \sim T^3$) and compressibility ($\kappa \sim T^2$) \cite{Yang}, making
this phase easily discernable by means of thermodynamics measurements.
Moreover, we argue that the chirality of any Weyl fermion is electrically switchable 
by reversing the ferroelectric polarization.

{\it Conclusions.---} In summary, we have considered the role of relativistic
SOC in the family of recently predicted hexagonal hyperferroelectrics, unveiling
an extremely rich physics that is strongly intertwined with ferroelectric
functionalities. Focusing our attention on two prototypical systems, namely
NaZnSb and KMgBi, we found  that electrically tunable bulk Rashba spin
splittings dominate the low-energy features of the bandstructure, putting forward HHs as
new FERSC whose cross-coupled functionalities should persist in the limit of
quasi-2D thin films. Increasing the strength of SOC, e.g, by replacing Sb with
Bi, may lead to a strong 3D topological insulating phase, with the concomitant
appearance of massless Dirac fermions at inequivalent surfaces. The Dirac cones
are found to be strongly modulated by the ferroelectric switching, opening
interesting perspectives, e.g., for domain engineering and control of
topological $p$-$n$ junctions. We further investigated the properties of
heterostructures where a few layers of HH KMgBi are interfaced with a normal (and
topologically trivial) ferroelectric; while the hyperferroelectric nature of
KMgBi is confirmed by persistent polar distortions even in the limit of few
atomic layers, we found that changing the thickness of the HH layer allows for a
tuning of the bulk band-gap and for the appearance of topological interface states.
Finally, we showed that a Weyl semi-metal phase can be achieved by alloying
KMgBi in a diluite solution with a topologically trivial hyperferroelectric.

{\it Note.---} During the completion of the paper, we became aware of a theoretical 
study of Rashba effects in the class of hexagonal ferroelectrics \cite{Narayan}.
On the other hand, in the present work we focus on the interplay between
hyperferroelectricity and relativistic effects induced by SOC, including
Rashba effect, 3D topological insulating and Weyl semimetal phases.

We acknowledge CINECA for providing us computational resources and Henry cluster
from North Carolina State University. D.D.S.~and A.S.~acknowledge the
CARIPLO Foundation through the MAGISTER project Rif.~2013-0726.
D.V.~acknowledges ONR Grant N00014-12-1-1035. The authors
acknowledge A.~Narayan for useful correspondence. 

\bibliography{biblio}

\end{document}